\documentclass[prx,twocolumn,preprintnumbers,superscriptaddress,longbibliography, 10pt]{revtex4-2} 

\usepackage{graphicx}
\usepackage{bm}
\usepackage{amsmath}
\usepackage{amssymb}
\usepackage{amsthm}
\usepackage{amsfonts}
\usepackage{slashed}
\usepackage{latexsym,epsfig,bbm}
\usepackage{algorithm}
\usepackage{algorithmic}
\usepackage{thmtools}
\usepackage{thm-restate}

\newcommand{\bra}[1]{\langle{#1}|}
\newcommand{\ket}[1]{|{#1}\rangle}

\newcommand{\bopk}[3]{\langle{#1}|{#2}|{#3}\rangle}

\newcommand{\figref}[1]{Fig.~\ref{#1}}

\newcommand{\Tr}{\mathrm{Tr}}
\usepackage{color}
\definecolor{blue}{rgb}{0,0.2,1}

\definecolor{red}{rgb}{0.9,0,0}

\newcommand{\past}[1]{\overleftarrow{#1}}
\newcommand{\fut}[1]{\overrightarrow{#1}}
\newcommand{\pastfut}[1]{\overleftrightarrow{#1}}

\newtheorem{Lem}{Lemma}
\newtheorem{prop}{Proposition}

\newtheorem{Cor}{Corollary}

\newtheorem{Def}{Definition}

\newcommand{\prf}[1]{\begin{proof}#1\end{proof}}

\theoremstyle{plain}

\begin{document}

\title{Causal memory and structural complexity of correlated quantum sequences}

\author{Thomas J.~Elliott}
\email{physics@tjelliott.net}
\affiliation{Department of Physics \& Astronomy, University of Manchester, Manchester M13 9PL, United Kingdom}
\affiliation{Department of Mathematics, University of Manchester, Manchester M13 9PL, United Kingdom}
\affiliation{Centre for Quantum Science and Engineering, University of Manchester, Manchester M13 9PL, United Kingdom}

\date{\today}

\begin{abstract}
What makes one quantum process more complex than another? We adopt an operational approach inspired by classical complexity science, which characterises structure through the memory required for causal generation. We develop a fully-quantum extension of computational mechanics for quantum space series: Correlated sequences of quantum outputs that do not return to interact with their generator. For stationary, memoryful quantum space series admitting finite-dimensional, primitive causal models, we show that any memory-minimal causal generator can be chosen to produce outputs isometrically, without a discarded environment. The resulting canonical matrix product state representation uniquely determines the irreducible memory required for generation. In particular, the minimum memory entropy is exactly the entanglement entropy across the past-future bipartition, giving entanglement a direct operational interpretation as irreducible memory for causal generation. The same structure determines the minimum memory dimension through the corresponding Schmidt rank. Within this class, we further show that the fundamental entropy rate vanishes, that causal generation is symmetric under time reversal, and that all memoryful processes possess infinite quantum Markov order with respect to arbitrary sharp measurements. These results establish an operational framework for quantifying intrinsic quantum structure, revealing deep connections between causal generation, complexity, memory, entropy, and entanglement.
\end{abstract}
\maketitle 

\section{Introduction}

It is often said that quantum systems are complex. Why do we think this? One may point to the promises of quantum computing~\cite{nielsen2000quantum}, or to non-classical phenomena such as coherence~\cite{streltsov2017colloquium}, entanglement~\cite{horodecki2009quantum}, discord~\cite{modi2012classical}, and magic~\cite{haug2023stabilizer}. Here, we approach the question through the information required to support a system's behaviour: What information must a quantum system retain to causally generate a correlated sequence of quantum outputs?

In the classical setting, this question lies at the heart of \emph{Computational Mechanics}~\cite{crutchfield1989inferring, shalizi2001computational, crutchfield2012between}, which identifies structural complexity in a dynamical system or stochastic process with the minimum memory required to predict its future behaviour. This captures a notion of intrinsic computation in the dynamics~\cite{crutchfield1994calculi}: A process possesses structure when information about its past must be retained to generate its future.

Over the last decade or so, efforts have been made to extend computational mechanics to the quantum regime~\cite{gu2012quantum, mahoney2016occam, elliott2018superior, aghamohammadi2018extreme, binder2018practical, loomis2019strong, liu2019optimal, elliott2022quantum, riechers2025identifiability}. In particular, \emph{quantum} models can simulate the behaviour of \emph{classical} stochastic processes with lower memory overhead~\cite{gu2012quantum, monras2016quantum, aghamohammadi2017extreme, garner2017provably, elliott2020extreme, elliott2021memory, elliott2022quantum, wu2023implementing, fanizza2024quantum, yang2025dimension, sundar2026quantum, sundar2026dimension} and thermal dissipation~\cite{loomis2020thermal, thompson2025energetic} than possible with any classical model. Beyond offering prospective quantum advantages for stochastic simulation, this reveals that the memory required to represent complex structure depends on the physics available to the model. It also leads to phenomena such as the ambiguities of simplicity~\cite{aghamohammadi2016ambiguity, suen2017classical} and optimality~\cite{liu2019optimal, loomis2019strong}.

Conversely, computational mechanics has been applied to probe structure in quantum processes and systems~\cite{crutchfield2008intrinsic, wiesner2008computation, tan2014towards, suen2022surveying, venegas2020measurement, venegas2023optimality, gier2025intrinsic}. Much of this work takes a semiclassical approach, studying the computational mechanics of a classical stochastic process extracted from observations of a quantum system~\footnote{A notable exception is the work of Gier and Crutchfield~\cite{gier2025intrinsic}, which considers sequences of unmeasured quantum states output by a classical hidden Markov model controller.}. Such observation-based processes can possess considerable classical complexity even when generated by a small quantum system, such as a qubit undergoing repeated measurement by a single instrument~\cite{elliott2020extreme}. This raises a further, fundamental question: How much of this complexity belongs to the underlying quantum process, and how much arises from the chosen observations and their classical description?

To address this, we seek a \emph{fully-quantum} computational mechanics, where both the process generated and the memory of its model are simultaneously treated in their quantum formalism. Frustrating such an approach is the richness of quantum dynamics; amongst the difficulties, we have the freedom to choose between complementary measurement instruments~\cite{milz2021quantum}, the invasiveness of these measurements~\cite{taranto2023hidden}, and the distinction between quantum temporal and spatial correlations~\cite{fitzsimons2015quantum, song2024causal}. The process tensor, or quantum comb, framework provides a general means of describing such processes and continues to yield fruitful insight into their structure~\cite{lindblad1979non, chiribella2008quantum, pollock2018non, milz2021quantum, taranto2019quantum}.

To make progress in this endeavour, we focus on a restricted -- yet broad -- class of quantum processes, which we term \emph{quantum space series}. Conceptually, a quantum system emits a quantum output state at each discrete timestep. These outputs may be retained and measured individually or collectively, but are not returned to interact with the generating system. The resulting sequence thus possesses spatial quantum correlations between outputs emitted at different time points -- hence the name. This retains the freedom to choose between measurement instruments while excluding feedback from interventions on emitted outputs into the generating dynamics.

Quantum space series connect naturally with established descriptions of quantum processes. Finite-memory quantum space series constitute finitely-correlated states~\cite{fannes1992finitely}, which have previously been studied as quantum processes~\cite{fannes2010quantum}, and admit description in terms of matrix product density operators~\cite{verstraete2004matrix}. Connections between quantum memory and matrix product representations have previously been established for quantum models of classical stochastic processes, where the model memory cost coincides with the past-future entanglement of an associated quantum state~\cite{yang2018matrix}. Related notions of memory size and complexity have also been formulated through purified process-tensor representations of non-Markovian open quantum evolution~\cite{guo2022reconstructing}. Closely-related structures also arise in quantum Markov chains~\cite{gudder2008quantum, accardi2020quantum, godley2023adaptive} and sequential ancilla-mediated generation schemes such as collision models~\cite{ciccarello2022quantum}. Entropy and randomness in quantum dynamics have also been studied through quantum dynamical entropy constructions~\cite{alicki1994defining, alicki2001quantum, alicki2005quantum}, while related notions of entropy rate and excess entropy have recently been developed for Choi-state representations of process tensors~\cite{chang2026much}. These connections provide a setting in which to ask how the structure of an unmeasured quantum output sequence determines the memory required for its causal generation.

In direct analogy with the relationship between hidden Markov models and computational mechanics in the classical setting, we seek the irreducible memory of a faithful causal model of a quantum space series. Causality requires that any junk information discarded to the environment by the model supply no predictive information beyond that available from the measured past. We impose this across sharp measurement histories, with measurement choices playing a role analogous to inputs in classical input-output processes~\cite{barnett2015computational}. This makes the modelling requirement independent of a preferred measurement basis.

Our central result concerns stationary, memoryful quantum space series admitting finite-dimensional, primitive causal models. We allow general generators that retain a memory, emit an output, and discard an environment at each timestep. Despite this freedom, we show that a model minimising both statistical and topological memory costs can be chosen to generate the outputs isometrically, without a discarded environment. Thus, the matrix product state structure of the minimal model follows from causal memory minimisation. Its canonical form identifies the required memory, yielding an operational interpretation of structures familiar from tensor network theory. In particular, the quantum statistical complexity is precisely the entanglement entropy across the past-future cut: The correlations across this partition quantify the irreducible memory required for causal generation. The resulting framework connects causal generation, memory, and the informational architecture of quantum space series. Within the finite-dimensional, primitive causal class, we further establish a vanishing fundamental entropy rate and the absence of causal asymmetry. These conclusions concern the unmeasured quantum sequence; classical processes obtained by measuring its outputs may still exhibit extensive randomness and different memory requirements. We illustrate these distinctions through explicit examples.

This manuscript begins with an overview of the classical computational mechanics framework, and quantum extensions thus far, in Sec.~\ref{sec:framework}. We then introduce quantum space series in Sec.~\ref{sec:qss}, alongside their informational architecture and structural properties. Application of the developed framework to example processes is given in Sec.~\ref{sec:examples}, and we conclude with a discussion in Sec.~\ref{sec:discussion}.

\section{Framework}
\label{sec:framework}

Let us first recapitulate the notion of a classical stochastic process, such that we can better frame the exposition on our characterisation of structure in quantum processes with reference to how it appropriately generalises concepts from the classical theory. A discrete-event classical stochastic process~\cite{khintchine1934korrelationstheorie} is a collection of stochastic variables $X_t$ describing events $x_t\in\mathcal{X}$ labelled by a time $t\in\mathcal{T}$, governed by a joint distribution over all of $\mathcal{T}$. We call each $X_t$ an event, and denote a particular realisation by its lowercase counterpart $x_t$. We shall consider bi-infinite discrete-time processes with events at regular timesteps, such that $\mathcal{T}=\mathbb{Z}$, and denote a contiguous string of events $X_{t:t+L}:=X_t,X_{t+1},\ldots,X_{t+L-1}$, where the left index is inclusive, and the right exclusive. We can therefore express the full distribution for the stochastic process as $P(X_{-\infty:\infty})$, from which the probability of any finite length sequence can be deduced from the marginals. We will also restrict our focus to stationary, i.e., time-invariant, stochastic processes, such that $P(X_{0:L})=P(X_{n:n+L})\forall n\in\mathbb{Z},L\in\mathbb{Z}$. We can therefore without loss of generality take $t=0$ to be the present, and demarcate the process into a past $\past{X}:=X_{-\infty:0}$ and future $\fut{X}:=X_{0:\infty}$, with $\pastfut{X}:=X_{-\infty:\infty}$ denoting the full sequence.

We further introduce the concept of a model. A model is tasked with producing statistically-faithful samples from the future of the process, one timestep at a time, given any set of past observations~\cite{crutchfield1989inferring, crutchfield1994calculi, shalizi2001computational, crutchfield2012between}. To do this, a model must possess a memory system with states $m\in\mathcal{M}$, an encoding function $f:\past{\mathcal{X}}\to \mathcal{M}$ that determines the appropriate memory state given the past observations, and an update rule $\Lambda:\mathcal{M}\to\mathcal{M}\times\mathcal{X}$ that outputs the next event and updates the memory state accordingly; generally, $f$ and $\Lambda$ are stochastic maps. These can be represented by edge-emitting hidden Markov models (HMMs), where the nodes are the memory states, and the edge emissions the events. We call a model `faithful' if $P_\Lambda(\fut{X}|f(\past{x}))=P(\fut{X}|\past{x})\forall \past{x}$, and `causal' if the memory contains no information about the future that could not be determined from the past, i.e., $I(\fut{X};\mathcal{M}|\past{X})=0$, where $I(A;B)$ is the mutual information between variables $A$ and $B$~\footnote{The mutual information $I(A;B):=H(A)+H(B)-H(AB)$, where \mbox{$H(A):=-\sum_{a\in\mathcal{A}}P(a)\log_2[P(a)]$} is the Shannon entropy of $A$. The quantum equivalent is defined by replacing $H(A)$ with the von Neumann entropy $S(\rho):=-\Tr(\rho\log_2[\rho])$.}.

In classical complexity science, the theory of computational mechanics asserts that the structural complexity of such a stochastic process is given by the minimal possible memory required of a faithful, causal model of the process~\cite{crutchfield1989inferring, crutchfield2012between}. This minimal model can be systematically identified, and is referred to as the $\varepsilon$-machine of the process~\cite{shalizi2001computational}. The corresponding encoding function $f_\varepsilon$ maps pasts to the so-called causal states of the process $s\in\mathcal{S}$, satisfying the causal equivalence relation 
\begin{equation}
P(\fut{X}|\past{x})=P(\fut{X}|\past{x}')\Leftrightarrow f_\varepsilon(\past{x})=f_\varepsilon(\past{x}').
\end{equation}
There are two natural ways to parameterise the memory requirement, leading to two complementary measures of complexity: The statistical complexity $C_\mu:=H(\pi_s)$, where $\pi_s=\sum_{\past{x}\in s}P(\past{x})$ is the steady-state distribution of the memory, characterising the amount of information stored in memory; and the topological complexity $D_\mu=\log_2[|\mathcal{S}|]$ that quantifies the size (in bits) of the memory needed to store this information~\cite{crutchfield1989inferring}.

As noted above, while the $\varepsilon$-machine is the provably minimal faithful, causal classical model of a classical stochastic process, there exist quantum analogues of these models that can manifest the same process with a (sometimes substantially) lower memory requirement~\cite{gu2012quantum, mahoney2016occam, elliott2018superior,  aghamohammadi2018extreme, binder2018practical, loomis2019strong, liu2019optimal}. This implies that a simple quantum system, measured in a single, fixed basis, is capable of manifesting behaviour that can appear arbitrarily complex to a classical observer~\cite{elliott2020extreme, venegas2020measurement}. Thus, we need a means to characterise structure in quantum processes from a fully-quantum perspective. The quest to define quantum stochastic processes has led to the development of quantum process tensors (also known as quantum combs)~\cite{pollock2018non, milz2021quantum}, that take an operational approach and define an object that maps from the observer's interaction with a system to the observed output statistics. Yet process tensors do not explicitly discriminate between `fundamental' invasiveness from quantum measurement and other forms of `driven' intervention as might be found in a classical input-output stochastic process~\cite{barnett2015computational}. As a first foray into characterising structure in quantum processes, we would like to avoid the latter. It is then tempting to consider the measurement statistics for a single fixed basis, in the manner of a quantum HMM (QHMM)~\cite{monras2011hidden, monras2016quantum, adhikary2020expressiveness, elliott2021memory, adhikary2021quantum, zonnios2025quantum, riechers2025identifiability} -- and then perhaps build upon this into an input-dependent approach where the input prescribes the measurement basis~\cite{cabello2018optimal}. But this ultimately leads to classical data, and should be seen as a semiclassical approach.

\section{Quantum Space Series}
\label{sec:qss}

\subsection{Formalism}

\begin{figure}
\includegraphics[width=\linewidth]{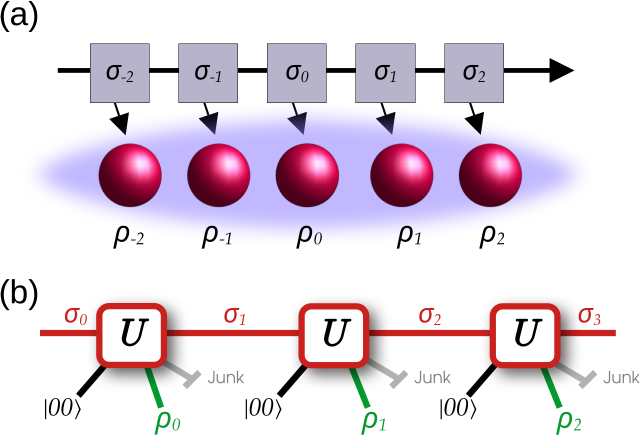}
\caption{{\bf Quantum space series.} (a) A quantum space series $\pastfut{\rho}$ consists of a one-dimensional array of spatially-correlated quantum systems, emitted by the time-invariant evolution of a persistent memory system $\sigma$. (b) This evolution can be represented by a unitary evolution $U$ acting on the memory and two blank ancillae at each timestep, with one of the resulting ancillae becoming the space series output for that timestep, and the other containing `junk' that is discarded into the environment.}
\label{fig:qss}
\end{figure}

Instead, we shall consider replacing the event alphabet $\mathcal{X}$ with quantum states $\rho^\mathcal{X}$, and define the size of the `quantum alphabet' $|\rho^\mathcal{X}|$ to be the dimension of the span of $\rho^\mathcal{X}$. Yet right away we hit a snag, which is that unlike classical processes, in general temporal quantum correlations can be distinguished from spatial quantum correlations. In a classical stochastic process, one can envisage that the observer is given a copy of the system state at each timestep, such that whatever they do with the state they receive the future dynamics are unaffected (in the sense of the Kolmogorov consistency condition~\cite{milz2020kolmogorov}). Thanks to the unclonability of a quantum state, this is not possible in the quantum setting, where the observer must be given the sole copy output quantum state to interact with. This leads to two scenarios: (a) the observer feeds the state back to the model after their interaction, potentially violating the Kolmogorov consistency condition, or (b) the observer keeps hold of the output state at each timestep. In the latter scenario, such a process can be seen as the observer receiving an infinite, homogeneous one-dimensional quantum chain state one site at a time, and the whole process (and any marginals) can be described by quantum systems spatially-distributed over the chain. We thus refer to such quantum processes as quantum space series [\figref{fig:qss}(a)]. These are the object of our study herein. 

In a quantum space series, we thus elevate the distribution of a classical stochastic process $P(\pastfut{X})$ to the state of a one-dimensional quantum chain $\pastfut{\rho}:=\rho_{-\infty:\infty}$, where the subscript labels the spatial position on the chain (acting as a proxy for the timestep), with contiguous strings denoted using analogous notation to the classical case. Marginals of the state correspond to marginals of the process, and we will consider stationary processes, such that $\rho_{0:L}=\rho_{n:n+L}\forall L\in\mathbb{Z},n\in\mathbb{Z}$. We can therefore without loss of generality denote 0 to be the present timestep, and correspondingly designate the past and future as $\past{\rho}:=\rho_{-\infty:0}$ and $\fut{\rho}:=\rho_{0:\infty}$ respectively. We also restrict our attention to processes generated by primitive models, as defined later. We remark that this notion of a quantum space series has close parallels with finitely-correlated states~\cite{fannes1992finitely}\footnote{Note however, that a finite-dimensional linear representation need not admit a finite-dimensional physical quantum realisation~\cite{fanizza2024quantum}}.

Recall that the Kolmogorov consistency condition for a (discrete-time) classical stochastic process demands that given a timestep $t$, the marginal distribution of the remaining timesteps (or any subset thereof) is identical when either (a) timestep $t$ is not observed, or (b) timestep $t$ is observed and the result forgotten~\cite{milz2020kolmogorov, milz2020when}. That is, $P(\pastfut{X}_{\backslash t})=\sum_{x_t}P(\pastfut{X}_{\backslash t}|x_t)P(x_t)$, where $\backslash t$ indicates an exclusion of timestep $t$ from the string. In other words, the marginals of the process are well-defined. For general quantum processes this condition does generally not hold, as not measuring is distinct from an unselective measurement, complicating the definition of appropriate marginals. However, this issue does not arise for quantum space series.\\
\begin{prop}
{Quantum space series satisfy the Kolmogorov consistency conditions.}
\end{prop}

\prf{Consider a general POVM $\{E^j\}$. The effect of an unselective measurement at timestep $t$ according to this POVM is to map the quantum space series state $\pastfut{\rho}$ to its marginal on the remaining sites $\sum_j\Tr_t(E^j_t\pastfut{\rho})$, where the subscript indicates that the measurement acts on the $t$ subsystem. Since partial traces are a linear operation, we can take the sum inside the trace, and use that $\sum_j E^j_t=\mathbb{I}$ to find that the unselective measurement leaves us in the marginal series $\Tr_t(\pastfut{\rho})=\pastfut{\rho}_{\backslash t}$, which is the same marginal as we would obtain performing no measurement. We can apply this repeatedly and iteratively through an arbitrary number of timesteps to see that this also holds when starting from any marginal of the full quantum space series. Thus, the Kolmogorov consistency condition holds; the marginals of a quantum space series are unaffected by whether or not measurements were made of other timesteps.}

The general form of a (quantum) model of a quantum space series must consist of a persistent memory system \textsf{M}, which at each timestep undergoes an interaction $\Lambda_q$ with an ancillary system resulting in an output drawn from $\rho^\mathcal{X}$, and an updated state of the memory system. The memory system must support a set of memory states $\sigma_\mathcal{M}$, where $\mathcal{M}$ is the set of memory state labels. The interaction $\Lambda_q:\sigma_\mathcal{M}\to\sigma_\mathcal{M}\times\rho^\mathcal{X}$ must be time-invariant, and represent a valid quantum channel (i.e., a CPTP map)~\cite{milz2021quantum}. The ancillary system must always be initialised in the same state, and we can without loss of generality consider its purified form and take this state to be $\ket{0}$. We can then invoke Stinespring's dilation theorem~\cite{stinespring1955positive} to cast $\Lambda_q$ as a unitary evolution $U$ that acts on the memory system and ancilla system to produce an updated memory state, an output state, and a `junk' state, the latter of which is discarded each timestep, playing the role of any (memoryless) stochastic influence from the environment [\figref{fig:qss}(b)]. Taking the ancilla to consist of two subsystems (i.e., $\ket{0}\ket{0}$), $U$ is specified in terms of its action on any set $\{\ket{j}\}$ of orthonormal basis states for the memory system:
\begin{equation}
\label{eq:genquantum}
U\ket{j}\ket{0}\ket{0}=\sum_{xyk} a_{xykj}\ket{k}\ket{x}\ket{y}.
\end{equation}
On the right-hand side, the first subsystem is the updated memory, the second the output, and the third the junk. $\{\ket{x}\}$ is an arbitrary orthonormal basis for the output system, and $\{\ket{y}\}$ for the junk. Tracing out the memory and junk subsystem leaves us with the corresponding output state, conditioned on the initial memory state (which may in general be in a superposition of the basis states $\{\ket{j}\}$, or a general mixture thereof). More specifically, given a general initial memory state $\sigma=\sum_mP(m)\ket{\sigma_m}\bra{\sigma_m}$, where $\ket{\sigma_m}:=\sum_jc_{mj}\ket{j}$, we have that
\begin{equation}
\rho^{X|\sigma}=\sum_{xx'}\sum_{mjj'ky}P(m)c_{mj}c_{mj'}^*a_{xykj}a_{x'ykj'}^*\ket{x}\bra{x'}.
\end{equation}
We can define a set of Kraus operators $\{A_{xy}\}$ that describe the update of the memory system; $\sigma\to\Phi(\sigma):=\sum_{xy}A_{xy}\sigma A_{xy}^\dagger$, where the $(k,j)$ element of $A_{xy}$ is given by $a_{xykj}$. We say that the model is primitive if this associated channel is primitive~\cite{godley2023adaptive}, i.e., there exists a positive, full rank state $\sigma_\pi$ such that $\Phi(\sigma_\pi)=\sigma_\pi$ and $\lim_{n\to\infty} |\Phi^n(\sigma)-\sigma_\pi|=0$ for all states $\sigma$ on the memory system.

Conditional forms of the map (i.e., conditioned on measurement of the output system) can be readily deduced. These Kraus operators can also be interpreted as site matrices of a (translationally-invariant) matrix product density operator (MPDO)~\cite{verstraete2004matrix}; this MPDO corresponds to the state embodied by the associated quantum space series generated by the model. We can also express the Kraus operators in different bases of the output system: Consider an orthonormal basis $\{\ket{\bar{x}}\}$ defined by $\ket{x}=\sum_{\bar{x}}u_{x\bar{x}}\ket{\bar{x}}$, where $u$ is a unitary matrix; the corresponding Kraus operators are given by $A_{\bar{x}y}=\sum_x u_{x\bar{x}}A_{xy}$. 

We can also define an associated encoding function for the memory. This encoding function is conditioned on the choice of probing of the quantum space series, essentially comprising of maps describing how the memory state is updated according to the relevant conditional form of $\Phi$. For example, if the system is unmeasured on a given timestep, $\sigma\to\Phi(\sigma)$; if a projective measurement is made with outcome $x$, $\sigma\to\sum_{y}A_{xy}\sigma A_{xy}^\dagger/\Tr(\sum_{y}A_{xy}\sigma A_{xy}^\dagger)$. For shorthand, let us denote the map associated with an outcome $x$ following a measurement $\Pi$ as $\Phi_{x,\Pi}$, and let $\Phi_{x_{t:t+L},\Pi_{t:t+L}}$ denote the concatenation of such maps over a series of consecutive timesteps. Then, we can define an encoding function $f_q:\past{\mathcal{X}}\times\{\past{\Pi}\}\to\sigma_{\mathcal{M}}$ that describes how the memory should be encoded to faithfully produce the correct future series given a set of past observations $\past{x}$ from measurements $\past{\Pi}$. Specifically, $f_q(\past{x},\past{\Pi})=\Phi_{\past{x},\past{\Pi}}(\sigma_\pi)$, where $\sigma_\pi$ is the steady-state of $\Phi$.

Finally, we introduce the notation $\rho^{X^L|\sigma}=\mathcal{T}_L(\sigma)$ to denote the output sequence following $L$ applications of the model on an initial memory state $\sigma$. $\rho^{X^L}_\pi$ denotes the specific sequence when the initial memory state is the steady state $\sigma_\pi$.

\subsection{Informational Architecture}

Let us now consider the informatic properties~\cite{crutchfield2003regularities} of a quantum space series.  For classical stochastic processes, a key measure of randomness is its entropy rate $h_\mu$, quantifying the uncertainty in the next event given an asymptotically-long sequence of past observations, i.e., $h_\mu:=\lim_{L\to\infty}H(X_0|X_{-L:0})$. An analogous quantity, the von Neumann entropy rate, has been studied for separable quantum processes~\cite{gier2025intrinsic}. We define this for general quantum space series, and refer to it as the fundamental entropy rate.

\begin{Def}
The fundamental entropy rate of a quantum space series is defined as $h_q:=\lim_{L\to\infty}S(\rho_0|\rho_{-L:0})$.
\end{Def}

\begin{prop}
For a stationary quantum space series, $h_q$ is well-defined.
\end{prop}
\prf{Consider the $L$-length `myopic' approximation to $h_q$, given by $h_q^L:=S(\rho_0|\rho_{-L:0})$. From the (strong) subadditivity of entropy it follows that $h_q^L\geq h_q^{L+1}\forall L\in\mathbb{N}$. Moreover, from the Araki-Lieb inequality~\cite{araki1970entropy}, it follows that $S(\rho_0)\geq|h_q^L|\forall L\in\mathbb{N}$. Thus, the sequence $h_q^L$ monotonically decreases with $L$, and has a lower bound; therefore, $h_q^L$ converges. Since $h_q=\lim_{L\to\infty}h_q^L$, the fundamental entropy rate is thus well-defined as the convergent limit of this sequence.}

An alternative, equivalent expression of the entropy rate $h_\mu$ for classical processes can be given in terms of the block entropy per symbol for sequences of sufficiently long length~\cite{crutchfield2003regularities}. Specifically, $h_\mu=\lim_{L\to\infty}H(X_{0:L})/L$. A similar expression holds for quantum space series, and can also be shown to be equivalent to the fundamental entropy rate. Let us define $\tilde{h}_q^L:=S(\rho_{0:L})/L$. We have that
\begin{align}
\sum_{l=L}^{L+N-1}h_q^l&=S(\rho_{0:L+N})-S(\rho_{0:L})\nonumber\\
&=N\tilde{h}_q^{L+N}+L(\tilde{h}_q^{L+N}-\tilde{h}_q^L).
\end{align}
The second term in the final expression is bounded by $LS(\rho_0)$, while for sufficiently large $L$, $h_q^L\approx h_q$, and hence the original sum is approximately given by $Nh_q$. Thus, if the quantum alphabet is of finite rank, for $N\gg L$ we obtain $h_q\approx\tilde{h}_q^{L+N}$, and hence $\lim_{L\to\infty} \tilde{h}^L_q=h_q$. 

We also remark on two further properties of the block entropy and entropy rate of a quantum space series. Firstly, it follows from $h_q^L\geq h_q^{L+1}$ that the block entropies are a concave function of $L$: $2S(\rho_{0:L})-S(\rho_{0:L-L'})-S(\rho_{0:L+L'})\geq0\forall L\in\mathbb{N},L'\in\mathbb{N}_{<L}$. Secondly, the reverse fundamental entropy rate $\past{h}_q:=\lim_{L\to\infty}S(\rho_{0}|\rho_{1:L+1})$ is equal to the fundamental entropy rate $h_q$, due to the stationarity of the sequences; the same holds true for finite $L$ approximations to these quantities.

With respect to a model, we can also consider the entropy rate of the junk system. Let us denote the sequences of junk by the states $\psi_{L:L'}$, and the corresponding entropy rate (and finite length approximations) by $h_\psi$ $(h_\psi^L)$. \\
\begin{prop}
$h_q=h_\psi$ for any model with finite memory dimension.
\end{prop}
\prf{Consider the entropy of the steady-state of the memory; this must be time-independent. Since the evolution at each timestep is unitary, the combined memory-output-junk system after $L$ timesteps must also have the same entropy as this. 
\begin{equation}
S(\sigma_\pi)=S(\sigma\rho_{0:L}\psi_{0:L})\geq |S(\sigma_\pi)-S(\rho_{0:L}\psi_{0:L})|,
\end{equation}
where we have for notational brevity slightly abused notation in referring to the joint states of the memory-output-junk system. Thus, $2S(\sigma_\pi)\geq S(\rho_{0:L}\psi_{0:L})$. We also have that
\begin{equation}
S(\rho_{0:L}\psi_{0:L})\geq |S(\rho_{0:L})-S(\psi_{0:L})|=L|\tilde{h}_q^L-\tilde{h}_\psi^L|.
\end{equation}
Therefore, if $S(\sigma_\pi)$ is finite, we have that $\lim_{L\to\infty}|\tilde{h}_q^L-\tilde{h}_\psi^L|=0$, and therefore $h_q=h_\psi$.}

We have termed $h_q$ the fundamental entropy rate as it pertains to the output sequence prior to any measurement. We can also consider measurement-dependent entropy rates: $h_{q,\Pi}^L=S(\rho_0|\Pi(\rho_{-L:0}))$, where $\Pi(\rho)$ indicates the distributions of outcomes of a measurement $\Pi$ of $\rho$.\\
\begin{Def}
{The observable entropy rate of a quantum space series with respect to a set of measurements $\mathcal{O}$ is defined as $h_{q,\mathcal{O}}:=\mathrm{inf}_{\Pi\in\mathcal{O}}\lim_{L\to\infty}h_{q,\Pi}^L$; its myopic approximation is defined $h^L_{q,\mathcal{O}}:=\mathrm{inf}_{\Pi\in\mathcal{O}}h^L_{q,\Pi}$}.
\end{Def}
There are multiple natural choices for $\mathcal{O}$ that lead to different observable entropy rates, each with different physical interpretations. For example, whether $\Pi$ is local or global with respect to the sites, whether or not $\Pi$ is a projective measurement or general POVM, and in the case of local measurements, whether or not these are in a single basis, and whether or not they are adaptive. We leave a more detailed study of these different observable entropy rates to a future work. Moreover, through quantum discord~\cite{ollivier2001quantum, henderson2001classical, modi2012classical}  we can relate the fundamental and observable entropy rates~\footnote{Recall that the discord is given by $D_{A,\mathcal{O}}(A;B):=I(A;B)-J_{A,\mathcal{O}}(A;B)$, where $J_{A,\mathcal{O}}(A;B):=\max_{\Pi\in\mathcal{O}} (S(B)- S(B|\Pi(A)))$.}. Specifically, we have that $D_{\rho_{-L:0},\mathcal{O}}(\rho_{-L:0};\rho_0)=S(\rho_{-L:0})-S(\rho_{-L:1})+\inf_{\Pi\in\mathcal{O}}S(\rho_0|\Pi(\rho_{-L:0}))$. Therefore,
\begin{equation}
h^L_{q,\mathcal{O}}-h^L_q=D_{\rho_{-L:0},\mathcal{O}}(\rho_{-L:0};\rho_0)\quad\quad\forall L\in\mathbb{N}.
\end{equation}

The excess entropy of a stochastic process~\cite{crutchfield2003regularities} quantifies the difference between the asymptotic block entropy and the estimate one would make of this by multiplying the entropy rate by the sequence length. We can define an analogous quantity for quantum space series.\\
\begin{Def}
{The excess quantum entropy of a quantum space series is defined as $E_q:=\lim_{L\to\infty}S(\rho_{0:L})-Lh_q$}.
\end{Def}
As with the entropy rates, we can also consider finite $L$-length approximations $E_q^L$ of the excess quantum entropy. The classical excess entropy can be equivalently restated in terms of the cumulative deviation of myopic approximations to the entropy rate, and the mutual information between the past and future of the process. The corresponding quantum excess entropy and these equivalent expressions have previously been established for separable quantum processes~\cite{gier2025intrinsic}. We show below that these expressions also hold for general quantum space series.
\begin{prop}
{Equivalent expressions for the excess quantum entropy are $E_q=\sum_{l=0}^\infty\left(S(\rho_0|\rho_{-l:0})-h_q\right)$ and $E_q=I(\past{\rho};\fut{\rho})$}.
\end{prop}
\prf{The first expression follows from expanding out the block entropy as a conditional entropy. Specifically, we can express
\begin{align}
E_q^L&=S(\rho_{0:L})-Lh_q\nonumber\\
&=S(\rho_{L-1}|\rho_{0:L-1})-h_q+E_q^{L-1}.
\end{align}
Iteratively applying the same expansion to the latter term in this expression, and taking the limit $L\to\infty$, we obtain the first expression in the proposition. For the second expression, stationarity gives
\begin{align}
I(\rho_{-L:0};\rho_{0:N})
&=S(\rho_{0:L})+S(\rho_{0:N})-S(\rho_{0:N+L})\nonumber\\
&=S(\rho_{0:L})-\sum_{l=N}^{N+L-1}h_q^l.
\end{align}
Taking $N\to\infty$ at fixed $L$, each term in the sum approaches $h_q$, giving
\begin{align}
\lim_{N\to\infty}I(\rho_{-L:0};\rho_{0:N})
=S(\rho_{0:L})-Lh_q=E_q^L.
\end{align}
Finally, taking $L\to\infty$ yields $E_q=I(\past{\rho};\fut{\rho})$. Since mutual information is non-decreasing as either block is enlarged, these successive limits give the full past-future mutual information, including when it diverges.}

\subsection{Structural Complexity}

Let us now turn to the question of structural complexity in a quantum space series. Recall that the general form of a model of a quantum space series is given by Eq.~\eqref{eq:genquantum}; we shall assume the model is faithful to the process we wish to generate. To specify the structural complexity, we must determine the memory required by the minimal (memory) causal model of the quantum space series. 

In the classical case, causality excludes information about the future in the model memory beyond that available from the past~\cite{shalizi2001computational}. In extending this principle to quantum space series, we distinguish information retained in the quantum memory from additional predictive information exported into a discarded environment. We require that, conditioned on a sharp measurement history of the outputs, access to the discarded environment cannot provide further information about future outputs. This condition concerns the junk actually produced by the model, rather than an arbitrary reference purifying its current memory state.

The requirement applies jointly to the junk produced over arbitrarily long finite blocks, and for every choice of sharp (i.e., rank-one projective) measurement on the corresponding outputs. Requiring this for all measurement choices parallels the treatment of inputs in classical input-output processes~\cite{barnett2015computational, thompson2017using, elliott2022quantum}, while ensuring that no particular measurement basis is privileged.

This is captured formally by the following definition:

\begin{Def}
A model is causal if the following holds for all $n,L\in\mathbb{N}$. Starting from its stationary memory state $\sigma_\pi$, let $\mathsf{P}$ comprise the next $n$ outputs, $\mathsf{E}$ all junk emitted during these $n$ timesteps, and $\mathsf{F}$ the subsequent $L$ outputs. For every sharp measurement on $\mathsf{P}$ and every outcome $x$ of non-zero probability,
\begin{equation}
\rho_{\mathsf{FE}|x}=\rho_{\mathsf{F}|x}\otimes\rho_{\mathsf{E}|x}.
\end{equation}
Note that the memory and any subsequently emitted junk are traced out in forming this state.
\end{Def}

This is a restriction on predictive information in the discarded environment; the absence of signalling from future to past alone does not imply it. For a causal model, even joint measurements on the discarded junk cannot improve predictions of the future conditioned on the specified output measurement history. A model with no junk satisfies this condition automatically.

Before stating our main theorem, we remark that there is a trivial class of quantum space series, which can be generated without need for any memory in the model. Such memoryless processes consist of independent and identically distributed sequences of individual quantum states, i.e., $\rho_{0:L}=\rho_0^{\otimes L}$ for all $L\in\mathbb{N}$. From hereon, we will consider only memoryful processes.

\begin{Def}
A quantum space series is memoryful if it is not an independent and identically distributed sequence of individual quantum states, i.e., $\rho_{0:L}\neq\rho_0^{\otimes L}$ for at least one $L\in\mathbb{N}$.
\end{Def}

\begin{Def}
The quantum statistical complexity $C_q$ and quantum topological complexity $D_q$ of a quantum space series are respectively defined as
\begin{align}
C_q:=&S(\sigma_\pi)\nonumber\\
D_q:=&\log_2[\mathrm{rank}(\sigma_\pi)],
\end{align}
{where $\sigma_\pi$ is the steady-state of the memory of the memory-minimal faithful, primitive causal quantum model, with minimality defined separately according to each of these measures}.
\end{Def}

\begin{restatable}{thm}{thmminmod}
\label{thm:minmodel}
{Every memoryful stationary quantum space series admitting a finite-dimensional, primitive causal model, has a memory-minimal model -- with respect to each of $C_q$ and $D_q$ simultaneously -- that requires no junk space. Its evolution can be specified by an isometry $V$ of the form}
\begin{equation}
\label{eq:uniquantum}
V\ket{j}=\sum_{xk}a_{xkj}\ket{k}\ket{x},
\end{equation}
where $\{\ket{j}\}$ is a set of orthonormal basis states for the memory system, $\{\ket{k}\}=\{\ket{j}\}$, and $\{\ket{x}\}$ is a set of orthonormal basis states for the output system. $V$ is unique up to change of basis for the memory and irrelevant global phase. $V$ can be realised by a unitary evolution $U$ satisfying $U\ket{j}\ket{0}=V\ket{j}\forall j$.
\end{restatable}

The proof is deferred to the Appendix. Intuitively, causality requires the junk and future outputs to be independent following every sharp measurement of an output block. For a memoryful stationary process, this forces the complete output and junk chains to factorise. The output chain is therefore a pure iMPS with a primitive canonical isometric generator. Its memory entropy and dimension attain lower bounds set by the past-future entanglement and Schmidt rank, respectively. Uniqueness follows from the output of this model being representable as an infinite matrix product state (iMPS)~\cite{perez2006matrix, cirac2021matrix, guta2015equivalence}.

The corresponding iMPS representation also provides a comparatively straightforward means of determining the structural complexity of a quantum space series: Construct such an iMPS representing the associated chain, and determine the Schmidt coefficients across any cut into two semi-infinite chains; the (base 2 logarithm) of the number of non-zero Schmidt coefficients corresponds to $D_q$, while $C_q$ is found from the Shannon entropy of the squares of these coefficients. For processes in this class, these memory costs are minimal among all faithful stationary finite-memory models, whether or not causality is imposed.

\begin{Cor}
A memoryful stationary quantum space series admits a finite-dimensional primitive causal model if and only if it has a finite-bond-dimension iMPS representation with primitive transfer channel.
\end{Cor}
\prf{The forward direction follows from Theorem 1. Conversely, the canonical isometric generator of such an iMPS is finite-dimensional, primitive, and causal because it produces no junk.}

Thus, within the memoryful stationary setting, the admissible processes are precisely the purely-generated primitive finitely-correlated states~\cite{fannes1992finitely}.

\subsection{Informational Architecture, Revisited}

With such constraints on causal quantum models, we can now also revisit certain informatic properties of quantum space series, and draw interesting conclusions. In the remainder of this Section we will consider quantum space series that have finite, non-zero $D_q$. Perhaps the most striking consequence is a constraint on the fundamental quantum entropy rate.\\
\begin{prop}
{For any quantum space series with finite $D_q$, $h_q=0$}.
\end{prop}
\prf{This follows from Proposition 3, and noting that in the absence of junk, as we may always have for the minimal causal quantum model, $h_\psi=0$.}

We remark that this result is already essentially known in the context of finitely-correlated states, where it has been shown that a `purely-generated' finitely-correlated state has zero entropy density~\cite{fannes1994finitely}.
 
The complement of this result is equally striking: Any quantum space series with non-zero entropy rate cannot be generated by a primitive causal model with finite memory dimension (unless the process is memoryless). If a quantum space series has non-zero entropy rate, any primitive finite-memory model that generates it `leaks' sideband information into the environment that could in principle be extracted to give greater predictive power as to future outputs in the series than could be obtained from observations of the past alone. In the computational mechanics literature this is referred to as `oracular' information~\cite{ellison2011information}. Proposition 5 also has implications for the excess quantum entropy.

\begin{Cor}{For a quantum space series with finite $D_q$, $E_q=\lim_{L\to\infty} S(\rho_{0:L})$}.
\end{Cor}
\prf{This follows directly from inserting $h_q=0$ into the definition of $E_q$. In other words, the entire asymptotic block entropy is excess quantum entropy.}
\begin{prop}{For a quantum space series with finite $D_q$, $C_q=E_q/2$}.
\end{prop}
\prf{Consider the memory-minimal, junk-free model prescribed by Theorem 1. The steady-state of the memory can be written in its eigenbasis $\{\ket{j}\}$ as $\sigma_\pi=\sum_jP_j\ket{j}\bra{j}$, where $\{P_j\}$ is the spectrum of $\sigma_\pi$. This can be purified to the state $\sum_j\sqrt{P_j}\ket{j}\ket{j}$. From this the second subspace can be used as the memory to generate the quantum space series up to length $L$, resulting in a state of the form $\sum_{j,x_{0:L}}\sqrt{P_jP(x_{0:L}|j)}\ket{j}\ket{x_{0:L}}\ket{\sigma_{\lambda(x_{0:L},j)}}$, where $\lambda(x_{0:L},j)$ defines the state in which the memory is left after starting in $\ket{j}$ and outputting $\ket{x_{0:L}}$ following sharp, local measurement of the output space in a prescribed basis, with $P(x_{0:L}|j)$ the probability of this outcome. Since this overall state is pure, the entropy of any bipartition is symmetric. Crucially, the joint entropy of the (purification of the) initial memory state and final memory state is thence the block entropy $S(\rho_{0:L})$. For $L\to\infty$, since the channel is primitive any correlation between the initial and final memory states vanishes asymptotically~\cite{yoshida2020asymptotic}. Therefore the joint entropy of the initial and final state asymptotically approaches the sum of their individual entropies -- which are both $C_q$. Thus, $C_q=\lim_{L\to\infty}S(\rho_{0:L})/2$. From Corollary 2, we then obtain $C_q=E_q/2$.}

Two further informatic properties of a stochastic process that we can export to quantum space series are the causal asymmetry~\cite{crutchfield2009time, thompson2018causal} and the Markov order~\cite{crutchfield2003regularities} of the process. Both of these exhibit strong constraints for quantum space series. The first of these, the causal asymmetry, quantifies the degree to which a process can be seen to have a preferred causal direction. Consider the time reversal of the stochastic process (such that the future takes the place of the past, and the most recent output in the past takes the place of the next output to be generated). The reverse statistical complexity $\past{C_\mu}$ is the statistical complexity of the time-reversed process, and the causal asymmetry is the difference between $C_\mu$ and $\past{C_\mu}$, i.e., how much additional information is required by the memory to causally model the process in one direction compared to the other~\cite{crutchfield2009time, thompson2018causal}. An analogous quantity can be defined for quantum space series, for the reverse quantum statistical complexity $\past{C_q}$.\\
\begin{prop}
{For any quantum space series with finite $D_q$, there is no quantum causal asymmetry; $C_q=\past{C_q}$}.
\end{prop}
\prf{The canonical iMPS representation can be generated sequentially in either direction. The reverse canonical generator is primitive and produces no junk, and is therefore causal. Applying Theorem 1 in each direction, the minimal memory spectra are the squared Schmidt coefficients across the same past-future cut. These coefficients are unchanged by exchanging the two sides, and hence $C_q=\past{C}_q$.}

The Markov order of a stochastic process specifies the relevant history length of the process with respect to the future~\cite{crutchfield2003regularities}. A memoryless process has Markov order 0. A Markovian process, where the future depends only on the most recent output, is Markov order 1. More generally, the Markov order is the smallest such length of the most recent past such that the remainder of the past and the future are conditionally independent. That is, the Markov order is the smallest $R$ such that $P(\fut{X}|X_{-R:0})=P(\fut{X}|\past{X})$; equivalently, the smallest $R$ such that $I(X_{-\infty:-R};\fut{X}|X_{-R:0})=0$. The notion of quantum Markov order has previously been defined for quantum stochastic processes expressed in terms of process tensors~\cite{taranto2019quantum,taranto2019structure}, where no non-Markovian process can have finite Markov order with respect to all possible instruments. For quantum space series, we define the Markov order operationally as the smallest $R$ such that, for every sharp projective measurement on the most recent $R$ outputs, the conditional future state depends only on the measurement and its outcome, independent of the earlier measurement history. This mirrors the classical definition of conditions on a classical measurement record, and is distinct from requiring vanishing quantum conditional mutual information in the unmeasured chain~\footnote{It is also distinct from the Markov length, as defined through the decay rate of quantum conditional mutual information~\cite{sang2025stability}.}.
\begin{prop}{For any memoryful quantum space series with finite $D_q$, the Markov order is infinite}.
\end{prop}

\prf{For a memoryless quantum space series, the Markov order is trivially 0. In the memoryful case, consider the memory-minimal model specified by Theorem 1, with $D_q>0$. Its injective canonical representation ensures that distinct memory states generate distinct complete futures. Moreover, for a sufficiently long output block, the corresponding outcome operators span the full space of memory operators. A suitable sharp measurement outcome can therefore realise an operator proportional to $\ket{v}\bra{w}$, for any chosen normalised memory states $\ket{v}$ and $\ket{w}$. Since $\sigma_\pi$ has full rank, conditioning on this outcome has non-zero probability and prepares $\ket{v}$. Thus, finite measurement histories can prepare pure states spanning the occupied memory space.

To be Markovian (i.e., Markov order 1), each possible outcome $x$ of any single-output sharp measurement must map all pure memory states prepared by earlier measurement histories into the same state $\ket{\sigma_x}$, whenever that outcome has non-zero probability. Since these states span the memory, this requires $A_x$ to be at most rank one. Denote each non-zero outcome operator as $A_x=\ket{u_x}\bra{v_x}$. For an outcome $\ket{\bar{x}}=\alpha\ket{x}+\alpha'\ket{x'}$ in another measurement basis, the corresponding operator is $A_{\bar{x}}=\alpha^*A_x+\alpha'^*A_{x'}$, which must also be at most rank one. If $\ket{u_x}$ and $\ket{u_{x'}}$ are linearly independent, this requires $\bra{v_x}$ and $\bra{v_{x'}}$ to be proportional; otherwise a combination with both coefficients nonzero has rank two.

Stationarity, $\sigma_\pi=\sum_x A_x\sigma_\pi A_x^\dagger$, and the full rank of $\sigma_\pi$ require the vectors $\{\ket{u_x}\}$ to span the memory. Since $D_q>0$, an independent pair therefore exists. Comparing any other outcome operator with this pair forces its input bra to be proportional to their common input bra: Otherwise its output vector would have to be parallel to both independent output vectors. Thus all $\bra{v_x}$ are proportional. This makes $\sum_x A_x^\dagger A_x$ have rank one, contradicting trace preservation. Therefore, Markov order 1 is impossible. The same argument applies to any finite Markov order $R$, replacing $A_x$ by the outcome operators of a joint sharp projective measurement on $R$ consecutive outputs. These operators likewise satisfy trace preservation and leave $\sigma_\pi$ stationary when outcomes are ignored. Thus, every memoryful quantum space series with finite $D_q$ has infinite Markov order.}

This means that no finite recent measurement window is guaranteed to determine the conditional future independently of the earlier measurement history for every choice of measurement. However, this does not rule out the possibility that there exists a specific finite length of measurement basis choices such that the outcomes of these measurements are guaranteed to allow one to specify the current memory state. The minimum possible such length can be taken to define a synchronisation order~\cite{crutchfield2003regularities} for a quantum space series. We leave an exploration of this to future work, but note that our first example in the following Section represents a quantum space series which can be synchronised with a single output when measured in the appropriate basis.

\section{Examples}
\label{sec:examples}

\subsection{Quantum perturbed coin}

Our first example is inspired by early works on quantum computational mechanics. The `perturbed coin' is a symmetric binary Markov chain with two outputs, $\{0,1\}$. The transition rule satisfies $P(x_t|x_{t-1})=p\delta_{x_tx_{t-1}}+(1-p)(1-\delta_{x_tx_{t-1}})$. That is, the same symbol will be emitted on two consecutive outputs with probability $p$. It was shown that a quantum model of this process can achieve an entropic memory reduction compared to the minimal causal classical model~\cite{gu2012quantum}. For our first space series, we will take the associated evolution of this quantum model as our generating isometry. When the outputs are measured only in the $\{\ket{0},\ket{1}\}$ basis, the process appears Markovian. However, measurements in other bases reveal memory effects.

The iMPS representation of the quantum perturbed coin can be constructed with the following pair of site matrices:
\begin{equation}
B_0=\begin{pmatrix}\sqrt{p} & \sqrt{1-p}\\ 0 & 0\end{pmatrix}\quad\quad\quad B_1=\begin{pmatrix} 0 & 0\\ \sqrt{1-p} & \sqrt{p}\end{pmatrix}.
\end{equation}

\begin{figure*}
\includegraphics[width=\linewidth]{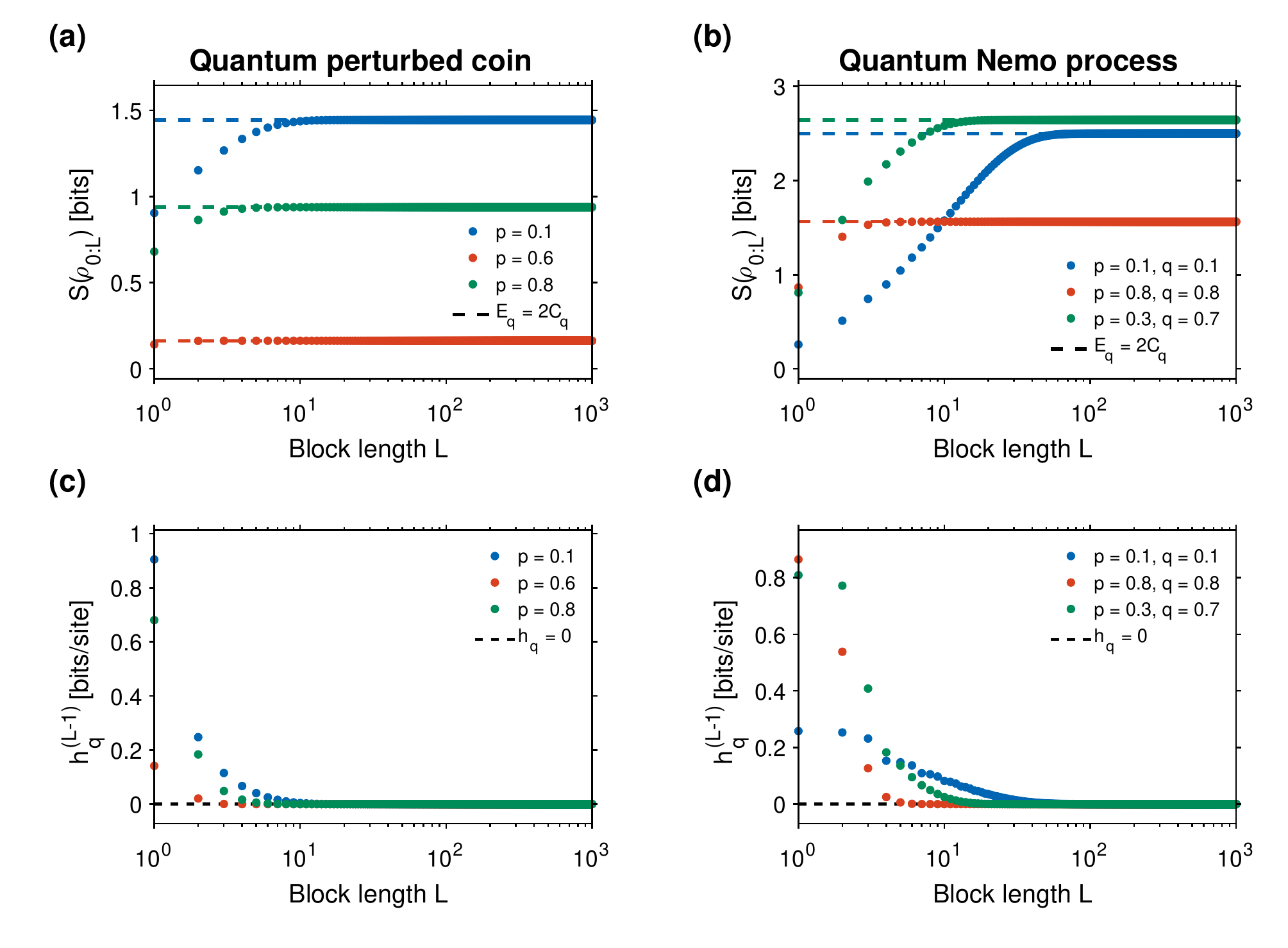}
\caption{{\bf Examples.} (a,b) Growth of quantum block entropy with increasing sequence length for the quantum perturbed coin and quantum Nemo processes at various parameter values, all showing convergence to the excess quantum entropy. (c,d) Analogous plots for the myopic fundamental entropy rate, all showing convergence to $h_q=0$.}
\label{fig:ex}
\end{figure*}

In \figref{fig:ex} we show (a) the growth of the block entropy $S(\rho_{0:L})$ with increasing sequence length $L$, and (c) myopic approximations of the fundamental entropy rate $h_q^L$, for parameters $p=0.1,$ 0.6, and 0.8. The block entropies converge to $E_q$, while the myopic entropy rates converge to $h_q=0$. For all parameter values, $|E^{1000}_q-2C_q|<10^{-12}$, and $|h_q^{999}|<10^{-15}$, suggesting that the values have essentially converged on their asymptotic quantities within 1000 timesteps. For all these parameters we have $C_\mu=1$ for the modelled classical process, while the full quantum space series (and the quantum model of the classical processes) have $C_q\approx(0.722, 0.081, 0.469)$ for the three parameters, indicating that the quantum statistical complexity of the space series is smaller than the classical statistical complexity of the classical process. Meanwhile, $E_q\approx(1.444, 0.163, 0.938)$ for the quantum space series, compared to the classical process' $E\approx(0.531, 0.029, 0.278)$, showing that measurement in the computational basis loses some of the past-future correlations present in the unmeasured quantum space series. We remark that while the space series has $h_q=0$ at all parameter values, the classical process has non-zero asymptotic randomness: $h_\mu\approx(0.469, 0.971, 0.722)$.

\subsection{Quantum Nemo process}

As a second example, we will consider another process inspired by a quantum model of a stochastic process. The Nemo process is a three-state renewal process, with binary alphabet $\{0,1\}$, and states $\{S_A,S_B,S_C\}$ with transition structure $P(S_A,1|S_A)=p$, $P(S_B,0|S_A)=1-p$, $P(S_C,0|S_B)=1$, $P(S_A,0|S_C)=1-q$, $P(S_A,1|S_C)=q$, and zero probability for all other transitions. For $0<p,q<1$, the classical process has infinite Markov order; while output 1 synchronises the system to $S_A$, an arbitrarily long finite string of 0 outputs need not do so. One possible quantum model of the process gives rise to a quantum sample state of the process given by an iMPS with site matrices given by:
\begin{equation}
B_0=\begin{pmatrix}0 & 0 & \sqrt{1-q}\\ \sqrt{1-p} & 0 & 0\\ 0 & 1 & 0\end{pmatrix}\quad\quad\quad B_1=\begin{pmatrix} \sqrt{p} & 0 & \sqrt{q} \\ 0 & 0 & 0 \\ 0 & 0 & 0\end{pmatrix}.
\end{equation}

In \figref{fig:ex} we show (b) the growth of the block entropy $S(\rho_{0:L})$ with increasing sequence length $L$, and (d) myopic approximations of the fundamental entropy rate $h_q^L$, for parameters $(p,q)=(0.1,0.1),$ (0.8,0.8), and (0.3,0.7). As with the quantum perturbed coin, the block entropies converge to $E_q$, while the myopic entropy rates converge to $h_q=0$; for all parameter values, $|E^{1000}_q-2C_q|<10^{-12}$, and $|h_q^{999}|<10^{-15}$, again suggesting that the values have essentially converged on their asymptotic quantities within 1000 timesteps. For the modelled classical Nemo process, $C_\mu\approx(1.58, 1.15, 1.56)$, while the full quantum space series (and the quantum model of the classical processes) have $C_q\approx(1.25, 0.781, 1.32)$ for the three parameter sets, again highlighting that the quantum statistical complexity of the space series is smaller than the classical statistical complexity of the classical process. Meanwhile, $E_q\approx(2.50,1.56,2.64)$ for the quantum space series, compared to the classical process' $E\approx(0.585, 0.499, 0.866)$, also showing that measurement in the computational basis loses some of the past-future correlations present in the unmeasured quantum space series. Again, while the space series has $h_q=0$ at all parameter values, the classical process has non-zero asymptotic randomness: $h_\mu\approx(0.318, 0.619, 0.624)$.

\section{Discussion} 
\label{sec:discussion}

We have introduced a framework for characterising structure in a class of intrinsically-quantum stochastic processes, which we term quantum space series. By extending the principles of computational mechanics into this fully-quantum setting, we identify the structural complexity of such a process with the minimal memory required for its faithful causal generation. The central result is that our causality condition forces any memoryful process with a finite-dimensional primitive causal model to admit a junk-free generator. Its canonical iMPS representation then attains the minimum memory entropy and dimension over all faithful stationary finite-memory models. This allows the complexity to be determined directly from the Schmidt spectrum across a past-future partition. In particular, the quantum statistical complexity is precisely equal to the corresponding entanglement entropy across this partition. This extends the connection between quantum model memory and entanglement~\cite{yang2018matrix} to the irreducible memory required for faithful causal
generation of the quantum space series itself.

The framework reveals a number of constraints on quantum space series admitting finite-dimensional, primitive causal models. Within this memoryful class, finite-dimensional causal memory requires vanishing fundamental entropy rate. Similarly, these quantum space series exhibit neither causal asymmetry nor finite quantum Markov order. Whilst these properties appear pathological when viewed through the lens of classical computational mechanics, they arise naturally once one accounts for the richer informational structure available in the quantum regime. In particular, a memoryful classical process with positive entropy rate, embedded as a quantum space series diagonal in a fixed product basis, admits no finite-dimensional primitive causal model. Such processes therefore lie outside the finite-dimensional class considered here, despite potentially admitting finite classical causal models. This distinction highlights the role of requiring causality for all sharp measurement choices, and motivates extensions in which the permitted observations are restricted.

The mathematical structures underlying quantum space series are closely-related to familiar constructions, including finitely-correlated states~\cite{fannes1992finitely, fannes1994finitely}, matrix product states~\cite{perez2006matrix}, and quantum Markov chains~\cite{gudder2008quantum, accardi2020quantum, godley2023adaptive}. The contribution of this work is to provide these structures with an operational and complexity-theoretic characterisation. This parallels the role of classical computational mechanics~\cite{crutchfield1989inferring, crutchfield1994calculi, shalizi2001computational, crutchfield2012between} within classical stochastic process theory: Not replacing existing descriptions, but identifying the information-processing architectures implicit within them.

The most immediate directions for future work concern the role of the process' observer. Here we focused on intrinsic properties of the process, independent of how it is interrogated. Yet the actions of observers influence the apparent complexity; meanwhile such an observer may be limited in their implementable measurements, such as being restricted to local probes, fixed measurement bases, finite-depth collective operations, or non-adaptive protocols. Such considerations naturally lead to observer-dependent notions of uncertainty, memory, and complexity. Understanding how these vary with observational capabilities may provide a bridge between intrinsic quantum structure and the semiclassical descriptions obtained from measurement statistics~\cite{tan2014towards, suen2022surveying, venegas2020measurement, venegas2023optimality}, and is key to the development of a fully-quantum computational mechanics. A further question is how the structural complexity can be inferred from finite observations. Spectral reconstruction methods for finitely-correlated states~\cite{fanizza2023learning} provide a natural starting point for this.

A second avenue is broadening the class of quantum processes under consideration. Quantum space series occupy one extreme of the spectrum of quantum stochastic processes, capturing purely spatial correlations whilst avoiding the complications associated with invasive temporal interventions. Extending the present framework to genuinely spatiotemporal quantum processes~\cite{milz2021quantum} is a vital next step to provide a fully-quantum theory of structural complexity applicable to arbitrary quantum dynamics. Such an extension would also connect with the question of when process memory must itself be quantum~\cite{giarmatzi2021witnessing}, and may also shed light on foundational questions concerning the distinction between spatial and temporal correlations~\cite{fitzsimons2015quantum, song2024causal} and the resources required to generate them. Finally, the framework developed here suggests a number of further connections between complexity science and quantum information theory. The relationship between causal generation, memory, and entropy raises the prospect of extending notions such as pattern manipulation~\cite{barnett2015computational, thompson2017using, elliott2022quantum} and work extraction~\cite{boyd2017leveraging, boyd2018thermodynamics, huang2023engines, thompson2025energetic} into a fully-quantum setting, to processes with quantum memory and entangled outputs.

\appendix

\section{Proof of Theorem 1}

Here we prove the main, underpinning result of this work, Theorem 1. We restate it here for convenience.\\
\thmminmod*
\prf{Recall that the general model of a quantum space series can be defined in terms of a unitary evolution on a memory system, output subspace, and junk subspace:
\begin{equation}
U\ket{j}\ket{0}\ket{0}=\sum_{xyk} a_{xykj}\ket{k}\ket{x}\ket{y}.
\end{equation} 
Equivalently, this can be represented as an isometry $V$ acting on the memory space, satisfying $V\ket{j}=U\ket{j}\ket{0}\ket{0}\forall j$.

We introduce a lemma required for the proof.

\begin{Lem}
Let $\rho_{\mathsf{ABC}}$ be a finite-dimensional tripartite state. Suppose that every non-zero-probability outcome of any sharp measurement on $\mathsf{A}$ leaves the combined state of $\mathsf{BC}$ as a product. Then, we must have that at least one of the following holds:
\begin{equation}
\rho_{\mathsf{ABC}}=\rho_{\mathsf{AB}}\otimes\rho_{\mathsf{C}}\quad\quad\mathrm{or}\quad\quad\rho_{\mathsf{ABC}}=\rho_{\mathsf{AC}}\otimes\rho_{\mathsf{B}}.
\end{equation}
\end{Lem}
\prf{Restrict $\mathsf{A}$ to the support of $\rho_\mathsf{A}$, of dimension $d$. If $d=1$, the statement trivially holds. For $d>1$, define the twisted state
\begin{equation}
\tilde{\rho}_{\mathsf{ABC}}=\frac{1}{d}(\rho_{\mathsf{A}}^{-\frac{1}{2}}\otimes I_{\mathsf{BC}})\rho_{\mathsf{ABC}}(\rho_{\mathsf{A}}^{-\frac{1}{2}}\otimes I_{\mathsf{BC}}).
\end{equation}
The marginal $\tilde{\rho}_{\mathsf{A}}=I_{\mathsf{A}}/d$; the inverse is taken only on the occupied support of $\mathsf{A}$. The twisting is invertible and local to $\mathsf{A}$. Conditioning the twisted state on a sharp measurement outcome $\ket{x}$ on $\mathsf{A}$ leads to the same normalised state on $\mathsf{BC}$ as would conditioning the original state on an outcome proportional to $\rho_{\mathsf{A}}^{-\frac{1}{2}}\ket{x}$. Thus, the conditional product structure is left unchanged by the twisting.

The normalised conditional twisted state on $\mathsf{BC}$ following measurement outcome $\ket{x}$ on $\mathsf{A}$ is $d\bopk{x}{\tilde{\rho}_{\mathsf{ABC}}}{x}$. Conditional expectation values on $\mathsf{BC}$ thus depend linearly on $\ket{x}\bra{x}$. Restrict $\ket{x}$ to an arbitrary two-dimensional subspace on the support of $\rho_\mathsf{A}$, and parameterise it with respect to the unit Bloch vector $\mathbf{n}$ of a qubit. Consider two arbitrary observables $O^B_{\mathsf{B}}$ and $O^C_{\mathsf{C}}$. Conditional expectations of these observables are affine functions of $\mathbf{n}$:
\begin{equation}
\langle O^B_{\mathsf{B}}\rangle_{\mathbf{n}}=b_0+\mathsf{b}.\mathsf{n}\quad\quad\mathrm{and}\quad\quad\langle O^C_{\mathsf{C}}\rangle_{\mathbf{n}}=c_0+\mathsf{c}.\mathsf{n}.
\end{equation}
Similarly, the joint conditional expectation $\langle O^B_{\mathsf{B}} O^C_{\mathsf{C}}\rangle_\mathbf{n}$ is also affine in $\mathsf{n}$. Conditional independence requires $\langle O^B_{\mathsf{B}} O^C_{\mathsf{C}}\rangle_\mathbf{n}=\langle O^B_{\mathsf{B}}\rangle\langle O^C_{\mathsf{C}}\rangle$. Thus, $(\mathsf{b}.\mathsf{n})(\mathsf{c}.\mathsf{n})$ must be constant on the unit sphere. Take $\mathbf{b}\neq 0$ and consider $\mathbf{b}\perp\mathbf{n}$, such that the constant is zero. It then follows that $\mathbf{c}$ must be 0, since $\mathsf{c}.\mathsf{n}$ vanishes whenever $\mathsf{b}.\mathsf{n}\neq 0$.

Thus, if any conditional expectation value of $O^B_{\mathsf{B}}$ varies over the two-dimensional subspace, the conditional expectation of every observable on $\mathsf{C}$ must be constant there, and hence the conditional state on $\mathsf{C}$ must be constant. Otherwise, by exchanging the subspaces in the above, we must have that the conditional state on $\mathsf{B}$ is constant. Hence, on each such subspace, at least one of the conditional marginals must be constant.

Suppose two different conditioning outcomes give different states on $\mathsf{B}$. It follows that the corresponding states on $\mathsf{C}$ must coincide. Any further outcome has a state on $\mathsf{B}$ differing from at least one of the first two, thus requiring that the state on $\mathsf{C}$ also coincides for this outcome. Extending this over all possible outcomes, all conditional states on $\mathsf{C}$ must be constant. If no such original pair exists for $\mathsf{B}$, then all conditional states on $\mathsf{B}$ are constant instead.

Let $\mathsf{C}$ be the subspace with the constant marginal. We have that $\bopk{x}{\tilde{\rho}_{\mathsf{ABC}}}{x}=\bopk{x}{\tilde{\rho}_{\mathsf{AB}}}{x}\otimes\tilde{\rho}_{\mathsf{C}}\forall \ket{x}$. Since the projectors $\ket{x}\bra{x}$ span the full Hermitian operator space, this requires $\tilde{\rho}_{\mathsf{ABC}}=\tilde{\rho}_{\mathsf{AB}}\otimes\tilde{\rho}_{\mathsf{C}}$. Taking $\mathsf{B}$ to be the constant marginal subspace instead gives $\tilde{\rho}_{\mathsf{ABC}}=\tilde{\rho}_{\mathsf{AC}}\otimes\tilde{\rho}_{\mathsf{B}}$. Undoing the twisting completes the lemma.}

We are now positioned to prove the main theorem. Start the model in its stationary memory state $\sigma_\pi$, and evolve it for $n+L$ timesteps. Let $\mathsf{P}$ comprise the first $n$ outputs, $\mathsf{E}$ the junk emitted during these timesteps, and $\mathsf{F}$ the subsequent $L$ outputs. Trace out the final memory and the junk emitted alongside $\mathsf{F}$. By causality, every sharp measurement on $\mathsf{P}$ gives $\rho_{\mathsf{FE}|x}=\rho_{\mathsf{F}|x}\otimes\rho_{\mathsf{E}|x}$ for each outcome of non-zero probability.

Applying Lemma 1, we have that the unmeasured state on $\mathsf{PFE}$ satisfies at least one of $\rho_{\mathsf{PFE}}=\rho_{\mathsf{PF}}\otimes\rho_{\mathsf{E}}$ or $\rho_{\mathsf{PFE}}=\rho_{\mathsf{PE}}\otimes\rho_{\mathsf{F}}$. Since the stationary process is memoryful, some finite blocks on opposite sides of a temporal cut are correlated. For sufficiently large $n$ and $L$, $\mathsf{P}$ and $\mathsf{F}$ contain such blocks, excluding the latter factorisation. Enlarging the windows, taking marginals, and using stationarity therefore establishes independence of the complete output and junk chains.

Consider now a dilation of the output space to include the junk, i.e., take a composite output space $\mathsf{Z}=\mathsf{XJ}$, as produced by the original isometry $V$. Since the memory channel is primitive, this combined
stationary chain is a pure finite-bond-dimension iMPS~\cite{perez2006matrix}. Since this chain factorises into output and junk, each factor is itself a pure finite-bond-dimension iMPS. The output factor inherits stationarity and mixing from the joint chain, and therefore has a primitive canonical isometric generator $V_X$~\cite{schon2005sequential}. This generator produces no junk and is consequently causal.

Let $\{\lambda_j\}_{j=1}^r$ be the non-zero squared Schmidt coefficients of the output chain across a past-future cut. The stationary memory of $V_X$ has this spectrum, giving memory costs $H(\{\lambda_j\})$ and $\log_2 r$ with respect to information stored, and (log-)dimension, respectively. To establish minimality, consider any faithful stationary finite-memory model of the same output chain. For finite past and future blocks $\mathsf{P}$ and $\mathsf{F}$, the future is obtained by applying a channel to the memory $\mathsf{M}$ while leaving $\mathsf{P}$ untouched. Data processing and the Araki-Lieb inequality give
\begin{equation}
I(\mathsf{P};\mathsf{F})\leq I(\mathsf{P};\mathsf{M})\leq 2S(\sigma_\pi).
\end{equation}
Taking the past and future windows to infinity, their mutual information is $2H(\{\lambda_j\})$, since the output chain is pure. Thus, $S(\sigma_\pi)\geq H(\{\lambda_j\})$.

For the dimension bound, now let $\mathsf{P}$ and $\mathsf{F}$ denote the complete past and future. Purify the joint state of past and memory using a reference $\mathsf{R}$, and dilate the channel from memory to future isometrically, retaining its auxiliary system $\mathsf{G}$. The resulting pure state on $\mathsf{PRFG}$ has Schmidt rank $d=\mathrm{rank} (\sigma_\pi)$ across $\mathsf{PR}:\mathsf{FG}$. Since its marginal on $\mathsf{PF}$ is the pure output chain, it factorises as $\ket{\Psi}_{\mathsf{PF}}\otimes\ket{\chi}_{\mathsf{RG}}$. Consequently,
\begin{equation}
d=r\,\mathrm{rank}(\rho_{\mathsf{R}})\geq r.
\end{equation}
Hence $\log_2[\mathrm{rank}(\sigma_\pi)] \geq\log_2 r$.

The canonical generator $V_X$ attains both bounds simultaneously. Uniqueness of the primitive canonical iMPS representation~\cite{guta2015equivalence} gives uniqueness of this junk-free generator up to a unitary change of memory basis and an irrelevant global phase.}

\acknowledgments
This work was funded by the University of Manchester Dame Kathleen Ollerenshaw Fellowship. We thank many friends and colleagues for insightful discussions on this topic over the last several years, that have helped shape this work.

{\bf{Generative AI use.}} Much of this work, including ideation, all initial proofs and examples, and a near-complete draft of the manuscript, were done without use of any generative AI. The final stages of manuscript preparation were carried out with critical feedback from OpenAI and Anthropic models. Most notably, this includes a sharpening of the statements of formal results and proofs, including a significant restructure of the proof of Theorem 1.

\bibliography{ref}

\end{document}